\documentclass{article}
\pdfoutput=1
\usepackage{amssymb}
\usepackage{amsmath}
\usepackage{amsthm}
\usepackage{clrscode}
\usepackage{float}
\usepackage{graphicx}

\floatstyle{ruled}
\newfloat{algfig}{thp}{lop}
\floatname{algfig}{Listing}
\newcommand{\assign}{$\leftarrow$ }

\newtheorem{theorem}{Theorem}

\title{Modelling Correlated Mobility}
\author{Mikael Asplund, Simin Nadjm-Tehrani\\
Department of Computer and Information Science\\
Link{\"o}ping University\\
\emph{\{mikael.asplund,simin-nadjm.tehrani\}@liu.se}}
\begin{document}
\date{}
\maketitle

\begin{abstract}
  When nodes in a mobile network cluster together or move according to common external factors (e.g., cars that follow the road network), the resulting contact patterns become correlated. In this work we address the question of modelling such correlated mobility movements for the analysis of intermittently connected networks. We propose to use the concept of node colouring time to characterise dynamic node contact patterns. We analyse how this model compares to existing work, and demonstrate how to extract the relevant data from actual trace files. Moreover, we show how this information can be used to derive the latency distribution of DTN routing protocols. Our model achieves a very good fit to simulated results based on real vehicular mobility traces, whereas models which assumes independent contacts do not.
\end{abstract}

\section{Introduction}
\label{sec:introduction}

In delay- and disruption-tolerant networks messages are propagated
using a store-carry-forward mechanism.  Such networks can have
applications for disaster area management~\cite{asplund09srds},
vehicular networks~\cite{lu10}, and environmental
monitoring~\cite{lahde07}. These systems offer many challenges and
have been extensively studied by the research
community~\cite{altman10,resta11,spyropoulos08,zhang07}.

Recent results indicate that to the extent that delay-tolerant
networks will be found on a larger scale, they will be composed of
islands of connectivity, that is, some parts that are well-connected
and some parts that are sparse. This in turn implies correlated and
heterogeneous contact patterns~\cite{aschenbruck10,hossmann11}. Most
existing analytic delay performance models fail to capture such
scenarios, since they assume independent and identical node contacts
patterns. Moreover, although there are also more complex mobility
models which include correlation~\cite{ciullo11,garetto07}, it is not
obvious how one should go about to create such models from real
existing traces.

We propose to use the concept of a node colouring process to
characterise dynamic and heterogeneous contact patterns. We present a
hierarchy of contact modelling models and show that our approach to
modelling the system is more general (models a wider set of mobility
patterns) compared to approaches where node contacts are assumed to be
independent. Due to the high-level nature of the colouring
abstraction, it can be used to analytically study performance
properties of message dissemination. We demonstrate how to derive the
latency distribution functions (i.e., not just the average case),
which allows obtaining more detailed information such as how delivery
ratio depends on the time-to-live parameter. We evaluate our
theoretical model using synthetic and real-life traces and show that
with real mobility, correlation is high which means that our model
provides a much better fit compared to earlier work.

There are three main contributions in this paper. First, a novel and
yet simple approach of characterising dynamic contact patterns is
introduced. Second, a hierarchy of possible connectivity modelling
assumptions showing how our approach relates to other connectivity
models is presented. Third, a scheme for deriving the routing latency
distribution for complex heterogeneous mobility models is provided
together with an experimental evaluation and validation of our
model. We show that heterogeneous mobility can result in such a high
correlation of contacts that theoretical results based on independent
inter-contact times are no longer valid whereas a model based on
colouring time distributions are able to correctly predict the
information dissemination latency.

The rest of the paper is organised as
follows. Section~\ref{sec:systemmodel} describes the system model and
the basic assumptions we make. Section~\ref{sec:hierarchy} presents a
model hierarchy of different assumptions on contact
patterns. Section~\ref{sec:latency} describes how to derive the
routing latency distribution given knowledge of the colouring rate
distribution. This latter distribution is discussed in
Section~\ref{sec:colouring}, and we explain how it can be determined
from mobility traces. Section~\ref{sec:evaluation} contains the
experimental evaluation. Finally, Section~\ref{sec:relatedworks} gives
an overview of the related work and Section~\ref{sec:conclusions}
concludes the paper.

\section{System model}
\label{sec:systemmodel}
Consider a system composed of $n$ mobile nodes (some possibly
stationary). Nodes can communicate when they are in contact with each
other. A contact is defined by a start and an end time between which
two nodes have the potential to communicate with each other (through
the existence of a link layer connection). We focus on connection
patterns and ignore effects of queuing and contentions. Moreover,
since we are interested in intermittently connected networks, the time
taken to transmit a message is assumed negligible in relation to the
time taken to wait for new contacts to be established. We call this
assumption A.

\subsection{Mobility as a point process}

\begin{figure}[t]
  \centering
   \resizebox{9cm}{!}{\includegraphics{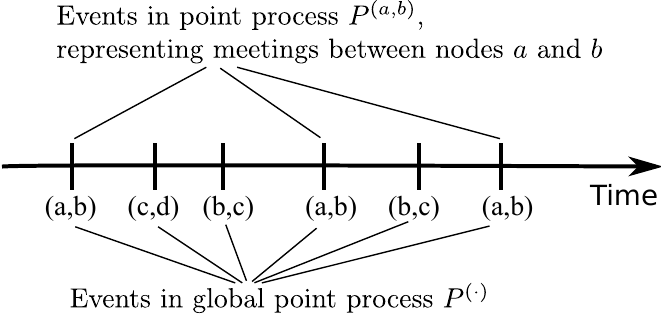}}
   \caption{Pairwise contact point processes}
  \label{fig:multivariate}
\end{figure}

Consider the sequence of contacts between some arbitrary pair of nodes
$a$ and $b$. This sequence of events can be mathematically represented
as a \emph{point process} on the time line, which we denote
$P^{(a,b)}$. By superimposing all such pairwise contact events we get
a global point process $P^{(\cdot)}$ where all contacts are
included. Figure~\ref{fig:multivariate} illustrates the general idea.

The pattern of the global point process corresponds to the node
meetings in the network, and thus is an abstract representation of the
node mobility. In this paper we aim to study how latencies of
delay-tolerant routing protocols can be predicted by analysing this
sequence of meeting events, and how the analysis is affected by
correlation between events. By making assumptions about the
characteristics of the contact point process, it is possible to create
mathematical models of message dissemination in the network. Since the
point process abstractly represents the node mobility, these
assumptions implicitly restrict the set of mobility models for which
the analysis holds. Therefore, it is desirable that the assumptions
are not stronger than necessary, and that they are as realistic as
possible.
\subsection{Colouring assumption}
\label{sec:colouringassumption}
Our approach to characterising the global contact point process is to
introduce a simple colouring process. This concept has been used
before to analyse routing latency~\cite{resta11,spyropoulos08}. The
novelty of our approach is to use properties of the colouring process
as an abstract \emph{representation of mobility}. Intuitively this
abstract representation captures the rate at which message
dissemination could take place in the network which is also closely
related to dynamic graph expansion~\cite{asplund12worst}.

\begin{figure}[tb]
  \centering
   \resizebox{9cm}{!}{\includegraphics{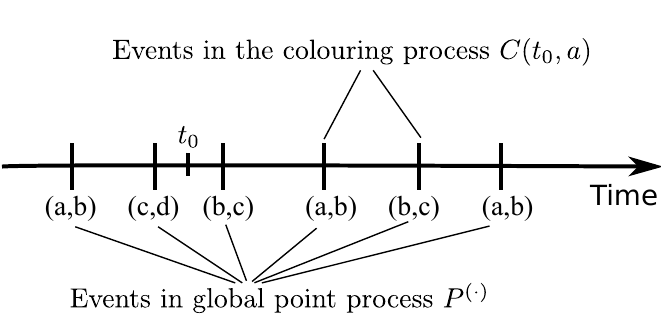}}
   \caption{Colouring process}
  \label{fig:colouring}
\end{figure}

The basic idea of the colouring process is that if some node $a$ is
coloured and subsequently comes in contact with node $b$, then node $b$
will also become coloured at the time of contact (if not already
coloured). Formally, a colouring process $C(t_0,o)$ for a given global
contact process is characterised by a start time $t_0$ and a source
node $o$ from which the colouring process originates (thus, $o$
becomes coloured at time $t_0$). Every contact event where a coloured
node comes in contact with an uncoloured node is a colouring
event. Figure~\ref{fig:colouring} shows a simple colouring process
$C(t_0,a)$ which starts at time $t_0$ by colouring node $a$. The
colouring events are the events where a coloured node meets an
uncoloured node, which in this case are when nodes $a$ and $b$, and
later $b$ and $c$ meet. Note that the first time $b$ and $c$ meet
after the time $t_0$, neither of the nodes are coloured, so this is
not a colouring event, whereas the second time they meet, node $b$ is
coloured so the second meeting with $c$ is therefore a colouring event.

The time between colouring events plays a crucial role in our model.
For a given contact point process, let $T_i$ denote the random
variable representing the time taken for a randomly chosen colouring
process to colour $i$ nodes, and let $f_i(t)$ and denote the
probability distribution function of $T_i$. Moreover, let $\Delta_i$
denote the random variable that describes the time taken for a
randomly chosen colouring process to colour \emph{one more node} given
that $i$ nodes are already coloured. We denote the probability
distribution function of $\Delta_i$ as $f_{\Delta i}(t)$. We can
express the time taken for a colouring process to reach $i+1$ nodes as
$T_{i+1} = T_{i} + \Delta_i$. Since the first node becomes coloured
immediately at the start of the colouring process, the time to colour
one node is $T_1 = 0$ and the time to colour two nodes is $T_2 =
\Delta_1$. 

The core assumption of our model which restricts the contact pattern
(and thereby constrains the mobility of the nodes) is that the incremental
colouring time should be independent from the colouring time so far.
More specifically, for any given colouring process, $T_i \bot \Delta_i$
for $2 \le i \le n-1$, where $\bot$ denotes the fact that two random
variables are independent (i.e., $X \bot Y \Leftrightarrow P(X\le x,Y
\le y) = P(X \le x)P(Y \le y)$). We call this assumption
$\bot_C$. Using this assumption together with the relationship
$T_{i+1} = T_{i} + \Delta_i$ we can characterise the probability
distribution function of $T_i$ as the convolution 
\begin{equation}
\label{eq:fi+1}
f_{i+1}(t) = (f_{i}*f_{\Delta i})(t)
\end{equation}

In the next section we compare the $\bot_C$ assumption with other
assumptions on node connectivity such as that of independent
inter-meeting times and show that our assumption is strictly weaker
(models a wider set of mobility patterns).

\section{Model hierarchy}
\label{sec:hierarchy}
Our goal in this paper is to mathematically analyse mobility patterns
with correlated meeting patterns. However, since we assume that
inter-colouring times are independent, it is interesting to analyse
how this assumption relates to the more common assumption of
independent contact times. The point of this analysis is to properly
understand these assumptions and the space of possible modelling
choices.

We start by describing a set of five assumptions (actually, the first
one is a non-assumption) on the independence of contacts. Each
assumption restricts the set of possible point processes (and thereby
which mobility patterns that can accurately be modelled with such a
point process).

\begin{itemize}
\item $\emptyset$ - No assumptions on the properties of the point processes (i.e., arbitrary mobility is allowed)\\
\\
Having no restricting assumptions on point processes allows all
possible contact (and thereby mobility) patterns. Unfortunately, this
makes it very hard to do any reasoning about system properties.

\item $\mathcal{S}$ - Every pairwise meeting process $P^{(a,b)}$ is
  \emph{stationary}.\\
  \\
  A stationary process is one where the intensity of events is
  constant over time. It does not mean that events arrive at a
  constant rate, but rather that the probability of finding a certain
  number of events in an interval of a given length is independent
  from the starting time of the interval. Note that a single stationary
  process cannot model intensity variations over a longer time period
  (e.g., rush hour and off peak times in a vehicular network). 

\item $\mathcal{R}$ - Every pairwise meeting process is a
  \emph{renewal process}: $X^{(a,b)}(i) \bot X^{(a,b)}(j)$ if $i \neq
  j$, where $X^{(a,b)}(i)$ denotes the time from event $i-1$ to event
  $i$ of the process $P^{(a,b)}$ counted after some origin time $t_0$.\\
\\
In a renewal process the time between two successive events is
independent from all other inter-event times. Note that in order for
the definition of $X^{(a,b)}(i)$ to be meaningful, the point
process needs to be stationary. A special case of the renewal process
is when the random variables $X^{(a,b)}(i)$ are exponentially
distributed, in which case the resulting point process is a Poisson
process.

\item $\bot_0$ - Non-overlapping pairs of nodes have independent meeting point processes, $P^{(a,b)} \bot P^{(c,d)}$ if $|\{a,b\} \cap \{c,d\}| = 0$\\
  \\
  This assumption basically states that if two node pairs are disjoint
  (i.e., there are four distinct nodes), then their respective
  contacts are also independent.

\item $\bot_{\le 1}$ - Two node-pairs of nodes with no more than one
  node in common have independent meeting point processes: $P^{(a,b)}
  \bot P^{(c,d)}$ if $|\{a,b\} \cap \{c,d\}| \le
  1$. \\
  \\
  This is similar to the previous assumption, but is stronger in the
  sense that any two node pairs with at most one node in common have
  independent meeting point processes. For example, node pairs $(a,b)$
  and $(b,c)$ are assumed to be independent from each other even if
  they have node $b$ in common.
\item $\bot_C$ - For any colouring process the time to colour one more
  node is independent from the time taken to colour $i$ nodes, $T_i
  \bot \Delta_i$ for $2 \le i \le n-1$.\\
  \\
  This is the assumption that we rely on in our analysis which was
  explained in Section~\ref{sec:colouringassumption}. Note that this
  assumption also requires stationarity of contacts (assumption
  $\mathcal{S}$), since we assume $T_i$ to be characterisable
  independently of when the colouring process starts.
\end{itemize}
\begin{figure}[tb]
  \centering
   \resizebox{12cm}{!}{
     \includegraphics{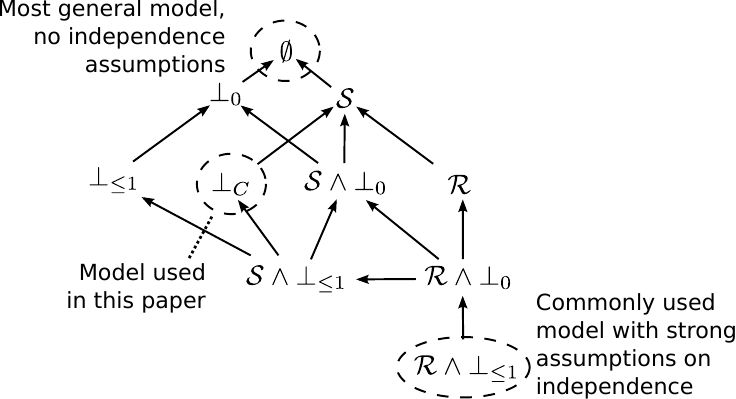}
   }
   \caption{Hierarchy of models characterised by assumptions}
  \label{fig:hierarchy}
\end{figure}

Based on these assumptions, we can create a hierarchy of models as
shown in Figure~\ref{fig:hierarchy}. The arrows in the figure denote
the relative strength of these assumptions. If $A \rightarrow B$, then
all behaviours that are modelled by $A$ are also modelled by $B$ (but
not vice versa). That is, $A$ has strictly stronger assumptions on the
contact patterns. It is possible to combine two assumptions by
assuming both of them to be true in the model (denoted with the
logical and symbol $\land$ in the figure).

At the top of the hierarchy is the most general model $\emptyset$
which makes no restrictions on node connectivity patterns. This allows
all possible contact patterns, including those that could never appear
in a wireless network with mobile nodes, but makes mathematical
analysis difficult. At the bottom of the hierarchy we find
$\mathcal{R}\land\bot_{\le1}$ which has been used extensively as a
model when studying intermittently connected
networks~\cite{altman10,resta11}. This gives tractable closed form
analytic expressions, but turns out to be too idealistic. Our model,
which is somewhere in between these two, provides a reasonable
trade-off between being able to analytically analyse the problem, and
still retaining a sufficient level of realism. We believe that it would
also be interesting to study the other models (especially those that
do not require stationarity) in the hierarchy as well, but it is
outside the scope of this paper. Some recent works have modelled
pairwise contacts as independent random variables but with
heterogeneous rates~\cite{picu12} which would add another dimension to
the lattice hierarchy.

We proceed with a brief explanation of how the hierarchy has been
derived. Many of the relationships between different independence
assumption models are immediate from the respective definitions. We
have already explained the fact that $\mathcal{R}$ is a stronger
assumption than $\mathcal{S}$, which we express as $\mathcal{R} \rightarrow
\mathcal{S}$. Moreover, the relation $\bot_{\le 1} \rightarrow \bot_0$
follows immediately from the respective definitions. Finally, when two
assumptions are combined, then the resulting assumption is also by
definition stronger than the individual assumptions, so for example
$(\mathcal{R}\land \bot_0) \rightarrow \mathcal{R}$.

The key relation which remains to show is $(\mathcal{S} \land
\bot_{\le 1}) \rightarrow \bot_C$, which basically says that the model
used in this paper ($\bot_C$) is strictly weaker (i.e., models a wider
set of mobility patterns) than the $\mathcal{S} \land \bot_{\le 1}$
model (as well as all stronger models since $\rightarrow$ is a
transitive relation). We present this proof in an appendix of the
paper. 

\section{Routing latency}
\label{sec:latency}

In this section we characterise the routing latency for epidemic and
multi-copy routing in intermittently connected networks. We use a
number of random variables to describe the colouring and routing
processes, Table~\ref{tab:notation} summarises the most important
notation. PDF is an abbreviation for probability density function and
CDF stands for cumulative density function, these abbreviations are
used throughout the paper. In this analysis we assume that the
distribution of the inter-colouring times $\Delta_i$ are known apriori
(we will later return to this assumption and discuss how to extract
this knowledge from existing mobility traces).
\begin{table}[tb]
\caption{Notation}
\label{tab:notation}
\center
\begin{tabular}{l p{47mm} @{\hspace{2mm}}|@{\hspace{2mm}} l p{47mm}}
  \hline
  \hline
  $n$ & Number of nodes in the system&
  $P(X)$ & Probability of  $X$ being true\\
  $T_i$ & Random variable, the time taken for a randomly chosen colouring process to colour $i$ nodes&
  $\Delta_i$ & Random variable, the time taken for a randomly chosen colouring process to colour one more node given $i$ coloured nodes\\
  $R$ & Random variable, the message delivery latency&
  $f_i(t)$ & PDF of the random variable $T_i$\\
  $f_{\Delta i}(t)$ & PDF of the random variable $\Delta_i$&
  $F_i(t)$ & CDF of the random variable $T_i$\\
  $F_{\Delta i}(t)$ & CDF of the random variable $\Delta_i$&
  $F_R(t)$ & CDF of the random variable $R$\\

  \hline
\end{tabular}
\end{table}

We start by deriving the routing latency for ideal epidemic
routing. This corresponds to the optimal performance any routing
algorithm can achieve. Thus, these results provide a useful
theoretical reference measure on what is good performance for a given
mobility model. Such a reference can also be of practical use to
decide whether the measured performance in some network is due to the
network characteristics or to the protocol implementation. We then
present an approximate model for the latency of multi-copy routing,
which uses the epidemic routing latency as a base.

\subsection{Latency of ideal epidemic routing}

Consider a randomly chosen time $t_0$, origin node $o$ and destination
node $d \neq o$. Let $R$ be the random variable that models the time
to route a message from $o$ to $d$ using ideal epidemic routing. We
will try to find the CDF of $R$, $F_R(t) = P(R \le t)$. Clearly, given
assumption A in Section~\ref{sec:systemmodel} (i.e., that the queuing and transmission times can be
neglected), this probability is the same as for $d$ being one of the
coloured nodes by the colouring process $C(t_0,o)$ after $t$ time
units.

Let $C_t$ be the random variable that models the number of coloured
nodes after $t$ time units. If $C_t = i$ then the probability that $d$
is coloured after $t$ time units is $(i-1)/(n-1)$ since if we remove
the source node $o$, there are $i-1$ coloured nodes and $n-1$ nodes in
total. Thus, we can express $F_R(t)$ as:
\begin{equation}
\label{eq:FR1}
F_R(t) = P(R \le t)= \sum_{i=1}^n P(C_t = i ) \cdot \frac{i-1}{n-1}
\end{equation}
Now let's consider the probability $P(C_t = i)$ that the number of
coloured nodes at time $t$ equals $i$. This is the same as the
probability that the time taken to reach $i$ nodes is less than or
equal to $t$ minus the probability that $i+1$ nodes can be reached in this time:
\begin{equation}
\label{eq:PC}
P(C_t = i) = P(T_i \le t) - P(T_{i+1} \le t)
\end{equation}
Combining equations~(\ref{eq:FR1}) and~(\ref{eq:PC}), and rewriting gives:
\begin{equation}
\label{eq:FR}
F_R(t) = \frac{1}{n-1} \sum_{i=2}^n F_i(t)
\end{equation}
where $F_i(t) = P(T_i \le t)$ is the CDF of the random variable $T_i$
which denotes the time for a randomly chosen colouring process to
reach $i$ nodes.  Recall that for the purpose of this analysis we
assume that the distribution of the inter-colouring times ($\Delta i$)
are known apriori. Moreover, from
Section~\ref{sec:colouringassumption}, we are able to iteratively
determine the PDF of $T_i$ through equation~(\ref{eq:fi+1}). The
conversion from $f_i(t)$ to $F_i(t)$ is just a matter of integrating
over time.

In summary, if we know the probability PDFs of the random variables
$\Delta_i$, we can use equation~(\ref{eq:fi+1}) to determine $f_i(t)$
(as well as $F_i(t)$). Equation~(\ref{eq:FR}) will then give us the
cumulative distribution function for the epidemic routing latency.
Listing~\ref{alg:getR} shows an algorithmic representation of how to
derive the distribution for $R$ using discrete distributions. The
procedure $\proc{conv}$ and $\proc{cumsum}$ are standard Matlab
functions and compute the convolution between two vectors and
cumulative vector sum respectively.

\begin{algfig}[t]
{\bf Input:} $f_{\Delta i}$ : Vector representing the PDF of $\Delta_i$
\begin{codebox}
  \li $f_2$ \assign $f_{\Delta 1}$
  \li \kw{for} i = 3 \ldots n
  \li \> $f_i$ \assign \proc{conv}($f_{i-1}$,$f_{\Delta i-1}$) \>\>\>\>\>\>\>\>/* equation~(\ref{eq:fi+1}) */
  \li $F_i$ \assign \proc{cumsum}($f_i$)  
  \li $F_R$ \assign $\frac{1}{n-1}\sum_{i=2}^n  F_i$ \>\>\>\>\>\>\>\>\> /* equation~(\ref{eq:FR}) */
  \li \kw{return} $F_R$
  \end{codebox}
  \caption{GetRoutingLatencyDistribution}
  \label{alg:getR}
\end{algfig}

Note that by knowing $R$ we can easily deduce the delivery ratio of a
protocol given a certain time-to-live (TTL) for each packet. The
probability that a message with TTL of $T$ will reach its destination
is simply $F_R(T)$ (i.e., the probability that the message will be
delivered within time $T$).

\subsection{Latency of multi-copy routing}
\label{sec:multicopymodel}
In the previous analysis we considered ideal epidemic routing, where
nodes propagate and multiply messages without discrimination. In real
networks, such a strategy results in severe congestion and poor
performance. The other extreme, when the system keeps a single copy of
the message at any point in time, can result in very long message
latencies unless additional knowledge can be used (e.g., geographic
forwarding). Between these two, there is a middle way where multiple
active copies are disseminated.  There are several ways of regulating
the number of active copies, including having the maximum number of
copies as a parameter, the Spray and Wait protocol by Spyropoulos et
al.~\cite{spyropoulos05} probably being the most well-known such
protocol.

In this section we describe how the colouring model of mobility can be
used also to approximate multi-copy routing latency. Our main
objective is not to extensively analyse multi-copy routing which has
been done commendably elsewhere~\cite{spyropoulos08multi}. Rather, our
goal is to demonstrate that the colouring model of correlated mobility
can also be used to analyse more advanced routing protocols than just
ideal epidemic routing. 

\begin{figure}[ht]
  \centering
  \includegraphics{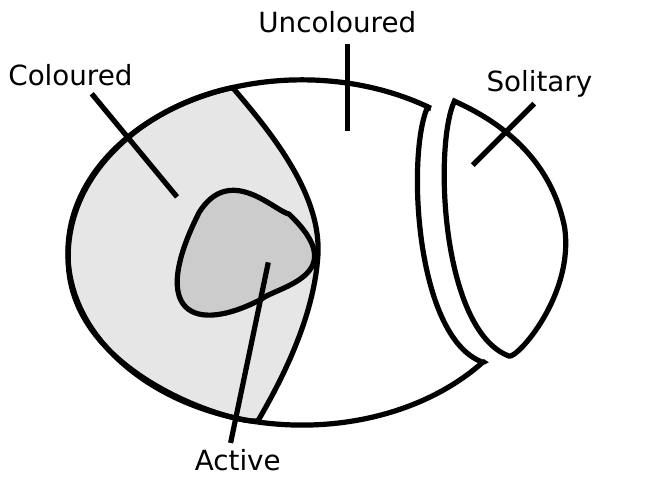}
  \caption{Possible node states in a multi-copy routing scenario with intermittent connectivity}
  \label{fig:snw}
\end{figure}

To make this approximation, we introduce two new node states as shown in
Figure~\ref{fig:snw}. In addition to a node being coloured or
uncoloured, we say that a coloured node can also be active, and that
an uncoloured node can be solitary. The basic intuition is that active
nodes are those which hold one of the message copies, and solitary
nodes are such nodes which seldom interact with any other nodes. 

The way we derive the final routing latency distribution is similar to
the process for epidemic routing. Thus, we first analyse how long it
would take a hypothetical colouring process to colour all the nodes,
and then use this information to derive the time taken to reach the a
given destination node. However, instead of directly using the
colouring distribution functions for $\Delta_i$, we consider a new
colouring process $C'(t_0,o,a)$ with inter colouring times $\Delta'_i$
where at most $a$ active nodes are allowed to colour other nodes. The
real-world interpretation is that a coloured node is a node such that
if it had been the destination of the message, then the message would
be delivered. So a node which meets an active node becomes coloured.

\begin{figure}[ht]
  \centering
  \resizebox{7cm}{!}{\includegraphics{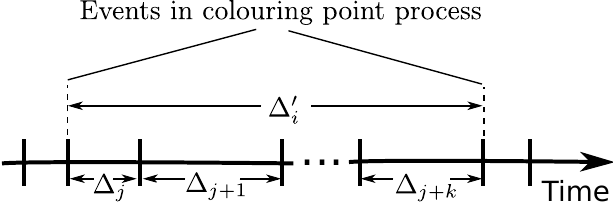}}
  \caption{The time between events in the colouring process $\Delta'_i$ equals the cumulative time between $k+1$ events in the super set process $\Delta_j + \Delta_{j+1} + \ldots + \Delta_{j+k}$  }
  \label{fig:multicopy}
\end{figure}

In order to determine the distribution function for $\Delta'_i$, we
take advantage of the fact while we do not know exactly the
characteristics of the point process $C'(t_0,o,a)$, we might know
properties of a point process containing a super set of the events in
$C'(t_0,o,a)$. Figure~\ref{fig:multicopy} illustrates the basic
idea. The time between colouring two events in this figure is the same 
as the time of $k$ events to happen in the super set process. In
general, the probability that $\Delta'_i$ is less than $t$ is (which
amounts to the cumulative distribution function $F_{\Delta' i}(t)$)
can be expressed as:

\begin{equation}
  \label{eq:superset}
  F_{\Delta' i}(t) = \sum_k P(i,k)F(i,k,t)
\end{equation}
where given $i$ coloured nodes, $P(i,k)$ represents the probability
that $k$ meetings of the super set process are needed for a colouring
event to occur, and $F(i,k,t)$ represents the probability that, given
$i$ coloured nodes, the time for $k$ events to occur in the super set
process is less than $t$. We will now proceed to find reasonable
approximations for $P(i,k)$ and $F(i,k,t)$. These approximations will
partly rely on assumptions of independence which will reduce the
fidelity of the model. However, by introducing these assumptions at a
late stage in the process we still retain important characteristics
about the node mobility. As a result, our approximate model still
provides good results in a scenario with correlated mobility as we
demonstrate in Section~\ref{sec:evaluation}.

Due to the nature of multi-copy routing we split the approximation of
$F_{\Delta' i}(t)$ in three phases. The first phase when there are
less than $a$ active nodes is similar to epidemic routing. In the
second phase $a$ active nodes is colouring the non-solitary nodes. In
the final phase the solitary nodes become coloured. Solitary nodes are
such nodes which seldom interact with any other nodes. Such behaviour
is seldom considered in homogeneous models, but can be seen in the
inter-colouring time distributions. If there are no solitary nodes,
then there is a symmetry $\Delta_i = \Delta_{n-i}$ (colouring the
first few nodes take equally long as colouring the last few nodes),
whereas if there are solitary nodes then $\Delta_i \ll \Delta_{n-i}$
for small values of $i$. That is, it takes much longer to colour the
last few nodes, than to colour the first few nodes. For the purpose of
this approximation we define the number of solitary nodes $s$ as the
largest $i$ so that $\Delta_i < 2\Delta_{n-i}$.

\begin{enumerate}
\item Initial phase: In this phase there are no more than $a$ coloured
  nodes ($1 \le i \le  a$), so the spreading process is essentially
  the same as epidemic routing. Therefore, we let $F_{\Delta' i} (t)=
  F_{\Delta i}(t)$
\item Middle phase: In this phase, there are more than $a$ coloured
  nodes, but the colouring has not yet reached the solitary nodes ($ a
  \le i \le (n-s)$). We assume that the likelihood of an active node
  meeting a solitary node is much less than the probability of meeting
  a non-solitary node. Moreover, if the chance of meeting a coloured
  node and uncoloured node is proportional to the number of respective
  nodes, the probability that $k$ meetings are needed to meet an
  uncoloured node is: $P(i,k) = \frac{n-s-i}{n-s-a}\left(
    \frac{i-a}{n-s-a} \right)^{k-1}$, where the first part of the
  expression is the probability of meeting an uncoloured node and the
  second part is the probability of $k-1$ consequtive meetings
  happening meeting only coloured nodes.  Finally, the probability
  that the time to meet $k$ nodes is less than $t$ can be approximated
  with $F(i,k,t) = (F_{\Delta a}(t))^k$, since there are $a$ nodes
  that actively colour other nodes and there need to be $k$ such
  colouring events.
\item End phase: In the final phase we assume that all non-solitary
  nodes have been coloured, and only the solitary nodes remain $(n-s)
  \le i \le (n-1)$. Considering the meetings between coloured and
  uncoloured nodes, the probability that a coloured node is active is
  $a/i$ so the probability that $k$ attempts are required is $P(i,k) =
  \frac{a}{i}\left( \frac{i-a}{i} \right)^{k-1}$. The CDF for the time
  of $k$ meetings between coloured and uncoloured nodes to occur can
  be expressed as $F(i,k,t) = (F_{\Delta i}(t))^k$. 
\end{enumerate}

These three steps together with equation~(\ref{eq:superset}), gives us
the cumulative distribution function $F_{\Delta' i}(t)$. By using
equation (\ref{eq:FR}), we can then calculate the latency distribution
function for multi-copy routing in the same way as we did for epidemic
routing.

\section{Colouring rate}
\label{sec:colouring}

Having derived the routing latency distribution based on knowledge of
the distribution of the incremental colouring time $\Delta_i$ we now
proceed to show how to find the distribution function $f_{\Delta i}(t)$
based on analysing existing mobility traces.

We consider two cases, when the mobility is homogeneous, and the more
interesting heterogeneous case. By homogeneous we mean that the
pairwise inter-contact times (i.e., the time between contacts) are
identical and independently distributed (often abbreviated iid). The
homogeneous case is not really novel in this context and is provided
here briefly in order to explain the baselines we have used and to
show that this case is also covered by our general approach.

\subsection{Homogeneous mobility}
\label{sec:homogeneous}
 
For the particular case of homogeneous mobility we adopt the model
$(\mathcal{R}\land\bot_{\le1})$ which states that all pair-wise
contact point processes should be stationary, renewal, and independent
(see Section~\ref{sec:hierarchy} for details).

Now consider a set of coloured nodes that wait for a new contact to
appear so that a new node can become coloured. The time they have to
wait is the smallest of all pairwise waiting times for all pairs where
one node in the pair is coloured and one node is uncoloured. If $i$
nodes are coloured, then there are $i(n-i)$ such pairs. Given the 
assumption of independence, we can express the CDF of $\Delta_i$ as:
\begin{equation}
\label{eq:homFi}
F_{\Delta i}(t) = P(\Delta_i  \le t) = 1-(1-F_\tau(t))^{i(n-i)}
\end{equation}
where $F_\tau(t)$ is the cumulative distribution of the
residual\footnote{The residual inter-contact time refers to the time
  left to the next contact from a randomly chosen time $t$, as opposed
  to the time to the next contact measured from the previous contact
  time.} inter-contact time between two nodes. We refer to Karagiannis
et al.~\cite{karagiannis10} for further explanation and how to derive
the residual distribution from the inter-contact distribution. If the
inter-contact time is exponentially distributed with rate $\lambda$,
then the residual waiting time is also exponentially distributed with
the same rate and the incremental colouring time $\Delta_i$ will be
exponentially distributed with rate $\lambda i (n-i)$.

\subsection{Heterogeneous mobility}
\label{sec:heterogeneous}

If node contacts are not independent, then deriving an expression for
the colouring distribution $\Delta_i$ will be more challenging. We
now proceed to present a first simple model for approximating it
from real heterogeneous traces.

In order to explain the rationale behind the model we  first show
some data from a real-life trace based on the movement of taxis in the
San Francisco area. The trace was collected by Piorkowski et
al.~\cite{piorkowski09} based on data made available by the
cabspotting project during May 2008 and we used a subset of the first
100 vehicles from the trace. In the simulation each taxi was assumed
to have a wifi device with a range of 550m.

\begin{figure}[tb]
  \centering
   \resizebox{8cm}{!}{\includegraphics{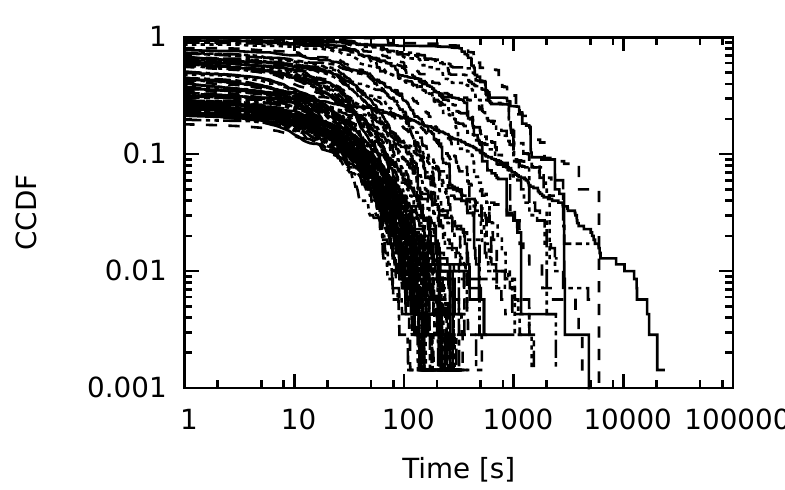}}
   \caption{Complementary Cumulative Distribution Functions (CCDF) of
     $\Delta_i$}
  \label{fig:ccdf}
\end{figure}

Fig.~\ref{fig:ccdf} shows the Complementary Cumulative Distribution
Function (CCDF) of each $\Delta_i$ (recall that $i$ corresponds to the
number of already coloured nodes) for the San Fransisco cab
scenario. We obtained this data by running 700 of colouring processes
on the contact trace and logging the time taken to colour the next
node. The plot uses a logarithmic scale on both axes to highlight the
characteristics of the distribution. This shows that they exhibit an
exponential decay (i.e., it approaches 0 fast, indicated by the sharp
drop of the curves.).

The second phenomena that we have observed is that due to clustering
of nodes, it is often the case that the next node can be coloured
without any waiting time at all. Based on these two basic principles
we conjecture that the colouring time can be modelled as either being
zero with a certain probability, or with a waiting time that is
exponentially distributed.

If $i$ nodes have been coloured, then we let $Con(i)$ denote the
probability that one of those $i$ nodes is in contact with an uncoloured
node (thereby allowing an immediate colouring of the next
node). Further we let $f_{\text{Exp}}(t,\lambda_i)$ denote the PDF of
the exponential distribution with rate $\lambda_i$. Then, we let the
PDF of the simple colouring distribution model be expressed as:
\begin{equation}
  \label{eq:colmodel}
  f_{\Delta i} =\begin{cases} Con(i) &\text{if } t=0\\
    (1-Con(i))*f_{\text{Exp}}(t,\lambda_i) &\text{otherwise}
    \end{cases}
\end{equation}

While this is clearly a simple model, it can be seen as a first step
towards modelling the colouring distribution and seems to work well
enough for the scenarios we have studied. We believe that further work
is needed to better understand how the colouring distribution is
affected by different mobility conditions. Note also that our general
scheme is not tied to this particular model and allows further
refinements.
    
\section{Evaluation}
\label{sec:evaluation}

To validate our routing latency esitmation model from Section~\ref{sec:latency} and to test whether it actually provides any
added value compared to existing models we performed a series of
simulation-based experiments. We used three different mobility models,
the random waypoint mobility model, a model based on a map of Helsinki
and a real-world trace from the cabs in the San Francisco area. After
explaining the experiment setup we give the details and results for
these models, first for epidemic routing, and then multi-copy
routing. Finally, we relate our findings on the effects of
heterogeneity for these cases.

We used the ONE Simulator~\cite{keranen09a} to empirically find the
ideal epidemic routing latency distribution and the routing latency of
Spray and Wait (which is an instance of a multi-copy routing
protocol) for the three different mobility models. For each mobility
model we ran the simulation 50 times. For the first 40000 seconds a
new message with random source and destination was sent every 50 to
100 seconds. The simulation length was sufficiently long for all
messages to be delivered. We used small messages of size 1 byte, and
channel bandwidth of 10Mb/s.

In addition to the simulated results we used two different
theoretical models to predict the latency distribution:

\begin{description}
\item[Colouring Rate:] This model uses equation~(\ref{eq:colmodel}) from
  Section~\ref{sec:heterogeneous} to model the colouring times. The
  necessary parameters $Con(i)$ and $\lambda_i$ are estimated from the
  trace file by sampling.
\item[Homogeneous:] This model assumes independent and
  exponentially\footnote{We also obtained nearly identical results
    when estimating the inter-contact distribution from the mobility
    trace, which we have excluded for lack of space.} distributed
  inter-contact times which are used to compute $f_{\Delta i}$ as
  described in Section~\ref{sec:homogeneous}. This has been a popular
  model for analysing properties of delay-tolerant
  networks~\cite{resta11,spyropoulos08,zhang07}.
\end{description}

In order not to get a biased value for the inter-contact time distributions
due to a too short sampling period, we analysed contacts from 200 000
seconds of simulation. To further reduce the effect of bias we use
Kaplan-Meier estimation as suggested by Zhang et
al.~\cite{zhang07}. 

\subsection{Epidemic routing}

{\bf Random Waypoint Mobility.} In order to validate our model against
already known results, we start with considering the random waypoint
mobility model. Despite its many
weaknesses~\cite{aschenbruck10,yoon03}, this model of mobility is
still very popular model for evaluating ad hoc communication protocols
and frameworks.  The network was composed of 60 nodes moving in an
area of $5km \times 5km$, each having a wireless range of $100m$. The
speed of nodes was constant $10m/s$ with no pause time.

\begin{figure}[htb]
  \centering
   \resizebox{8cm}{!}{\includegraphics{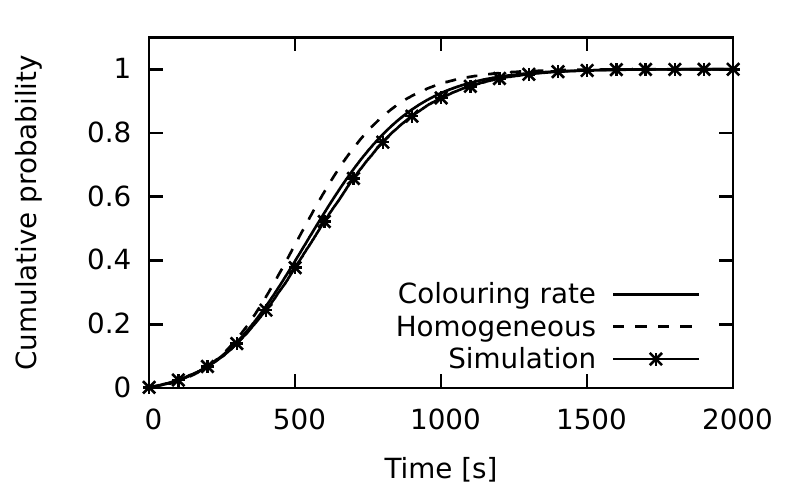}}
   \caption{Routing latency for random waypoint mobility}
  \label{fig:rwp}
\end{figure}

Fig.~\ref{fig:rwp} shows the results of the two theoretical models
and the simulation. The graph shows the cumulative probability
distribution (i.e., the probability that a message will has been
delivered within the time given on the x axis). As expected, both
models manage to predict the simulated results fairly well. In fact,
the exponential nature of the inter-contact times of RWP is well
understood and since the heterogeneous model is more general, we were
expecting similar results.

\begin{figure}[htb]
  \centering
   \resizebox{8cm}{!}{\includegraphics{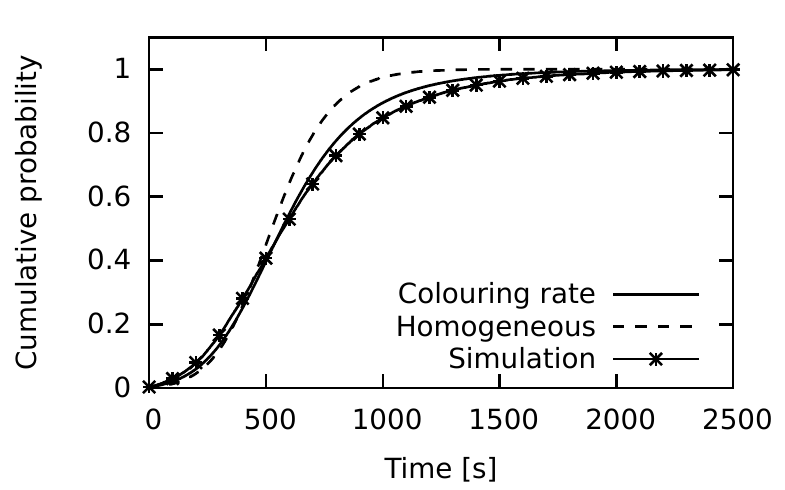}}
   \caption{Routing latency for Helsinki mobility}
  \label{fig:helsinki}
\end{figure}

{\bf Helsinki Mobility.} We now turn to a more realistic and
interesting mobility model, the Helsinki mobility model as introduced
by Ker{\"a}nen and Ott~\cite{keraenen07increasing}. The model is based
on movements in the Helsinki downtown area. The 126 nodes is a mix of
pedestrians, cars, and trams, and the move in the downtown Helsinki
area (4500x3400 m). We used a transmission range of 50 meters for all
devices.  Fig.~\ref{fig:helsinki} shows the results. Again both
theoretical models achieve reasonable results. However, due to the
partly heterogeneous nature of the mobility model, the homogeneous
model differs somewhat more from the simulated result. In particular,
we see that the s-shape is sharper compared to the observed
data. We further discuss possible explanations for this in
Section~\ref{sec:effects}.

\begin{figure}[htb]
  \centering
   \resizebox{8cm}{!}{\includegraphics{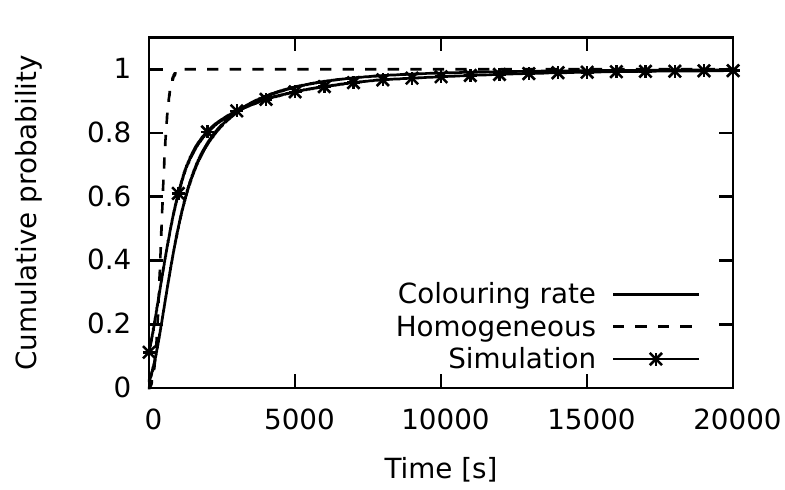}}
   \caption{Routing latency for San Francisco cab trace}
  \label{fig:cab}
\end{figure}

{\bf San Francisco Cabs.}
\label{sec:cab}
Finally, the last mobility trace we have analysed is a real-life trace
based on the movement of taxis in the San Francisco area as explained
in Section~\ref{sec:heterogeneous}. Fig.~\ref{fig:cab} shows the
results. In this case the homogeneous model fails to capture the
routing latency that can be observed in simulation. However, the
heterogeneous model based on equation~(\ref{eq:colmodel}) is still quite
accurate. We were surprised to find such a big difference between the
simulated data and the homogeneous model. Something is clearly very
different in this trace compared to the synthetic mobility models. An
estimate of the fraction of messages being delivered within an average
latency of 2500s in such a scenario would be misleadingly optimistic
by 20\%.

\subsection{Multi-copy routing}

\begin{figure}[ht]
  \centering
  \includegraphics{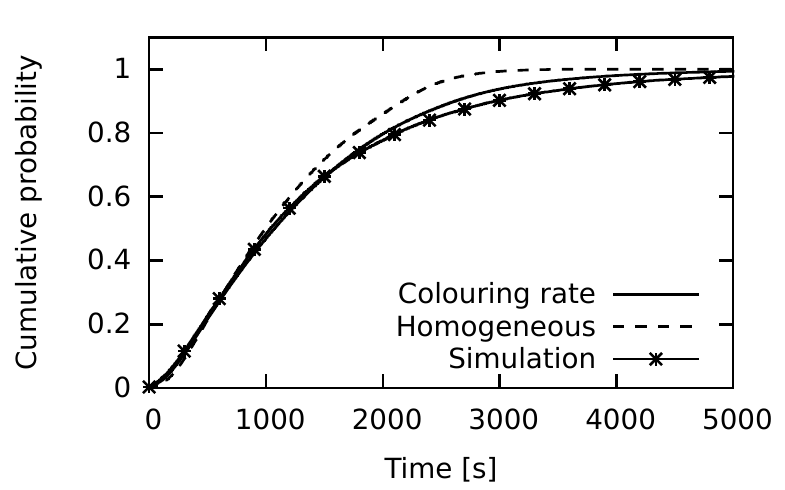}
  \caption{Spray and Wait with Helsinki Mobility}
  \label{fig:helsinki_snw}
\end{figure}

\begin{figure}[ht]
  \centering
  \includegraphics{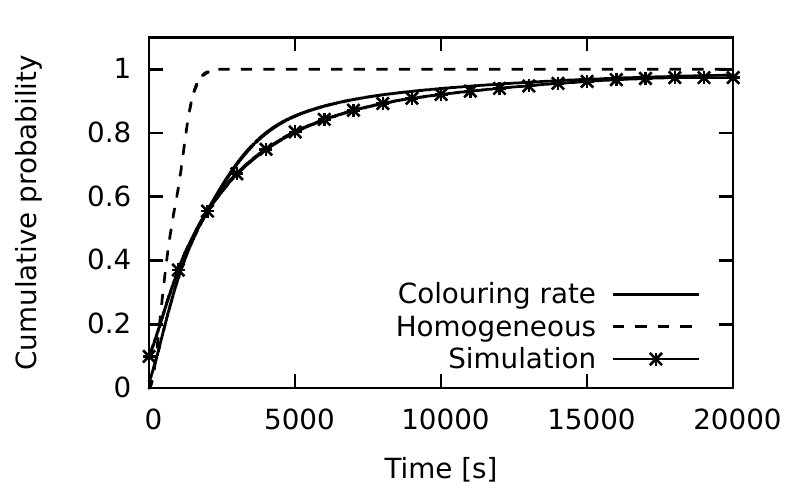}
  \caption{Spray and Wait with Cab Mobility}
  \label{fig:cab_snw}
\end{figure}

In Section~\ref{sec:multicopymodel} we proposed an approximate model
to predict the latency distribution of multi-copy routing, which was
based on our basic method of modelling correlated connectivity
patterns.  We now proceed to validate this model in the same manner as
for epidemic routing. We used a replication factor of 10, both in the
simulation and in calculations (i.e., the parameter $a$).  The results
from the Helsinki and San Francisco cabs mobility models are shown in
figures~\ref{fig:helsinki_snw} and~\ref{fig:cab_snw} (the results for
RWP all follow the simulated line closely so we omit this
curve). Clearly, the results are structurally similar to those of
ideal epidemic routing, but routing takes longer time (about twice the
time). This is expected since we exclude the effects of limited
bandwidth which otherwise significantly reduces the performance of
epidemic routing (due to congestion).

In terms of the ability of the theoretical model to predict the
results achieved in simulation, we can again observe that the model
which assumes homogeneous, independent mobility, works reasonably
well, for syntactic mobility, but not very well for the trace based on
real vehicular movements. However, when using a colouring-based
connectivity model the routing latency distribution closely matches
the simulated results, despite the fact our analysis of multi-copy
routing was only approximate.

\subsection{The effects of heterogeneity}
\label{sec:effects}

We have seen that the accuracy of the homogeneous model is high for
the random waypoint model, but is lower for the Helsinki model and
completely fails for the San Francisco cab trace. In this subsection
we present our investigation into why this is the case. We proceed by
identifying four different aspects of how this model differs from
reality.

{\bf Correlation.} We begin with the most striking fact of the
results presented so far. The homogeneous model is way off in
predicting the routing latency distribution in the San Francisco
case. There are a number of different ways that one can try to explain
this, but we believe that the most important one has to do with
correlation (i.e., non-independence) of events. The main assumption
that makes equation~(\ref{eq:homFi}) possible, and thereby the
homogeneous model is that the contacts between different pairs of node
are independent from each other. However, this seems to be a false
assumption.

\begin{figure}[tb]
  \centering
   \resizebox{8cm}{!}{\includegraphics{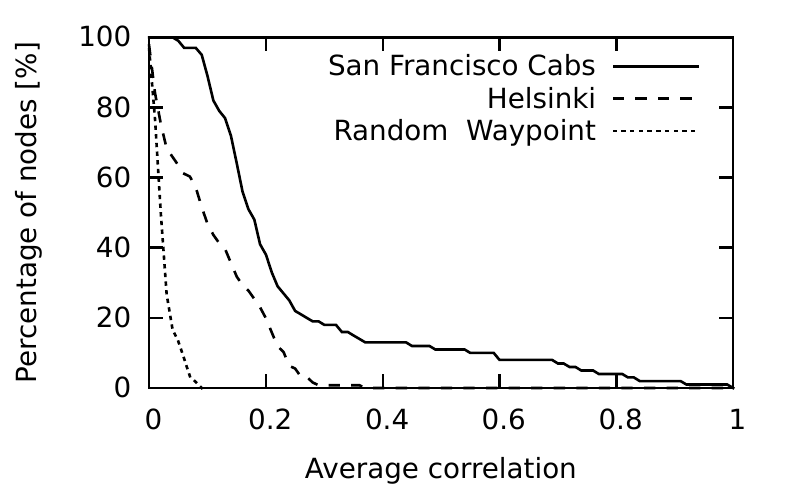}}
   \caption{Correlation of contacts}
  \label{fig:corr}
\end{figure}

We analysed the contact patterns of the three different mobility
models by considering the residual inter-contact times for each node
during a period of 20000 seconds.  Fig.~\ref{fig:corr} shows the
percentage of nodes for which the average correlation among their
contacts is higher than a given value (i.e., it is the complementary
CDF of nodes having a given average correlation). If the pairwise
contacts are independent, they will have no (or very low) average
correlation and we would expect to see a sharp decay of the curve in
the beginning of the graph. This is also what we see for the random
waypoint model. Since the nodes move around completely independently
from each other, the contacts also become independent. The Helsinki
trace shows a higher degree of correlation, but not as significant as
for the San Francisco cab case. In this case 40\% of the nodes have an
\emph{average} correlation of their contacts which is higher than 0.2
(a correlation of 1 would mean that all contacts are completely
synchronised). This shows a high degree of dependence and we believe
provides an explanation of the result we have seen in
Section~\ref{sec:cab}.

Note that correlated mobility does not necessarily lead to slower
message propagation, in fact there are results indicating the
contrary~\cite{ciullo11}. What we have seen is that the \emph{prediction} of
the latency becomes too optimistic when not taking correlation into
account. If the model assumes that contacts are ``evenly'' spread out
over time, whereas in reality they come in clusters, the results of
the model will not be accurate.

\begin{figure}[tb]
  \centering
   \resizebox{8cm}{!}{\includegraphics{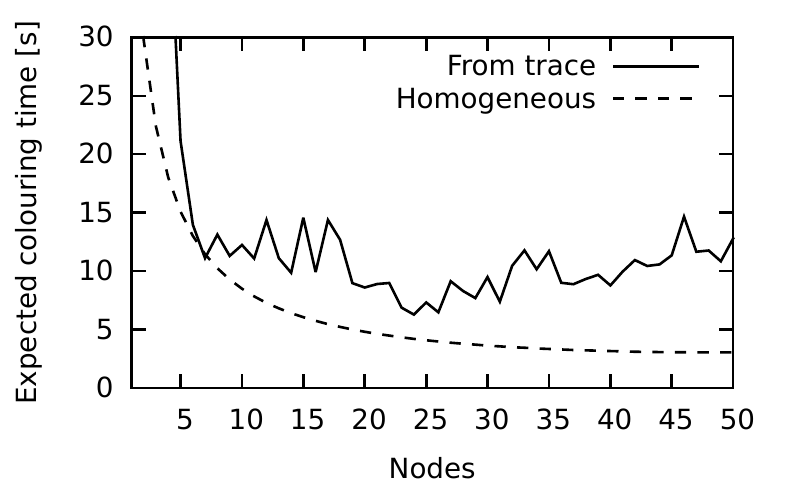}}
   \caption{Time to colour one more node in the San Francisco trace}
  \label{fig:cabcol}
\end{figure}

{\bf Lack of Expansion.} The second prominent effect is what we choose
to call lack of expansion (motivated by the close connection to
expander graphs~\cite{asplund12worst}). This means that the rate of
the colouring process seems not to correspond to the number of
coloured nodes. Fig.~\ref{fig:cabcol} shows the expected time to
colour one more node for the San Francisco trace. The x-axis
represents the number of nodes already coloured (up to half the number
of nodes). We can see that the homogeneous model predicts that the
time decreases (i.e., the rate of colouring increases) as the number
of coloured nodes increase. On the other hand, the data based on
sampling the distribution of $\Delta_i$ from the mobility trace file
(indicated as ``From trace'' in the figure) shows that after the first
5-10 nodes have been coloured, the rate is more or less independent
from the number of coloured nodes. We believe that this is partly due
to the fact that most of the node mobility is relatively local and
that nodes are often stationary for long periods of time.
 
\begin{figure}[htb]
  \centering
   \resizebox{8cm}{!}{\includegraphics{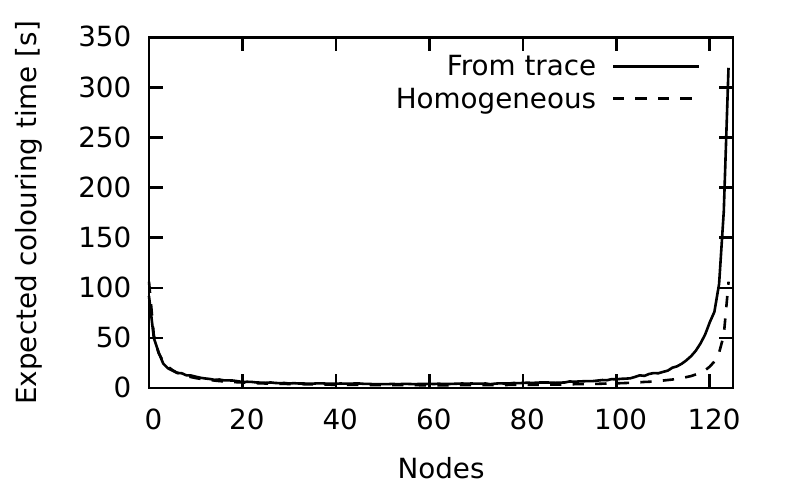}}
   \caption{Time to colour one more node with the Helsinki mobility model}
  \label{fig:helsinki-fdelta}
\end{figure}

{\bf Slow Finish.} Another effect that can be observed is that in some
rare cases it can take a very long time for a message to reach its
destination. For example in Fig.~\ref{fig:cab}, even after 10000
seconds not all messages have been delivered to their destinations. This
has to do with the fact that the time to colour all nodes is
significantly longer than to colour \emph{almost} all the nodes. The
models based on independent contacts predict that it takes the same
amount of time to colour the second node as it takes to colour the
last node. In both cases there are $n-1$ possible node pairs that can
meet and result in a colouring. However, we have seen that in reality
colouring the last node takes significantly longer (on
average). Fig.~\ref{fig:helsinki-fdelta} shows the effect for the
Helsinki trace, by plotting the expected colouring time as a function
of the number of coloured nodes. While the homogeneous model is
completely symmetrical around the middle, the actual data shows that
it takes roughly three times longer to reach the last node than to
reach the second node.

{\bf Fast Start.} Finally, we consider why the homogeneous model
predict a lower probability for delivering messages fast. This can be
seen in both the Helsinki and San Francisco cases, but is more
distinct in the former case. It can be seen visually in
Fig.~\ref{fig:helsinki} in that the homogeneous model has a slightly
flatter start compared to the other curves. This is because there is a
chance that when a message is created, the node at which it is created
has a number of neighbours. Thus, the message will not need to wait
any time at all before being transmitted. Or if we express it as a
colouring process, the time to colour the second node is sometimes
zero. For a model based on inter-contact times, this is not
considered.

\begin{figure}[tb]
  \centering
   \resizebox{8cm}{!}{\includegraphics{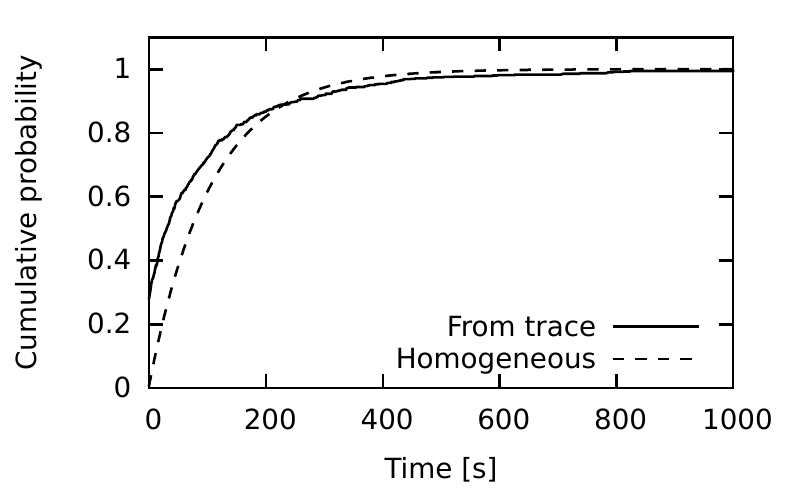}}
   \caption{CDF of the time to colour the second node}
  \label{fig:colsecond}
\end{figure}

Fig.~\ref{fig:colsecond} shows the CDF of $T_2$, (i.e., the time
taken to colour the second node) for the Helsinki case with the
colouring rate and homogeneous model. We see that both curves are
similar (the expected value for $T_2$ is the same for both models) but
that the start value differs. That is, in the homogeneous model, it is
predicted that the chance that the second node is immediately
coloured is zero, whereas in fact it is roughly 0.3. Recall that the
colouring time only reflects the contact patterns of the mobility and
does not consider message transmission delays.

In this section we have seen how heterogeneous mobility causes
correlated contacts and how that affects predictions of routing
latency. Our model which is based on colouring rate of nodes was the
only model able to accurately predict the routing latency distribution
in these cases.

\section{Related works}
\label{sec:relatedworks}

There is a rich body of work discussing detailed analytic models for
latency and delivery ratio in delay-tolerant networks. The work ranges
from experimentally grounded papers aiming to find models and
frameworks that fit to observed data to more abstract models dealing
with asymptotic bounds on information propagation. Many of these
approaches are based on or inspired by epidemiological
models~\cite{khelil02}. We have previously characterised the
worst-case latency of broadcast for such networks using expander graph
techniques~\cite{asplund12worst}. In a preliminary version of this
paper we developed the basic framework for deriving the latency of
ideal epidemic routing~\cite{asplund12analysing}. This paper extends
the latter work by incorporating an analysis of multi-copy routing as
well as presenting a more rigrorous matematical basis with a hierarchy
of connectivity models.

Closest to our work in this paper is that of Resta and
Santi~\cite{resta11}, where the authors present an analytical
framework for predicting routing performance in delay-tolerant
networks. The authors analyse epidemic and two-hops routing using a
colouring process under similar assumptions as in our paper. The main
difference is that our work considers heterogeneous node mobility
(including correlated inter-contact times), whereas the work by Resta and
Santi assumes independent exponential inter-contact times.

Zhang et al.~\cite{zhang07} analyse epidemic routing taking into
account more factors such as limited buffer space and
signalling. Their model is based on differential equations also
assuming independent exponentially distributed inter-contact times. A
similar technique is used by Altman et al.~\cite{altman10}, and
extended to deal with multiple classes of mobility movements by
Spyropoulos et al.~\cite{spyropoulos09}.

Kuiper and Nadjm-Tehrani~\cite{kuiper12framework} present a quite
different approach for analysing performance of geographic routing in
large scale. Their framework can be used based on abstract mobility
and protocol models together with extracting distributions for
arbitrary mobility models and protocols from simulation data (in
smaller scale). The main application area for this model is geographic
routing where waiting and forwarding are naturally the two modes of
operation in routing.

The assumption of exponential inter-contact times was first challenged by
Chaintreau et al.~\cite{chaintreau07} who observed a power law of the
distribution for a set of real mobility traces (i.e., meaning that
there is a relatively high likelihood of very long inter-contact
times). Later work by Karagiannis et al.~\cite{karagiannis10} as well
as Zhu et al.~\cite{zhu10} showed that the power law applied only for
a part of the distributions and that from a certain time point, the
exponential model better explains the data. Pasarella and
Conti~\cite{passarella11} present a model suggesting that an aggregate
power law distribution can in fact be the result of pairs with
different but still independent exponentially distributed
contacts. Such heterogeneous but still independent contact patterns
have also been analysed in terms of delay performance by Lee and
Eun~\cite{lee10}.

Our work on the other hand, suggests that the exact characteristic of
the inter-contact distribution is less relevant when contacts are not
independent. Correlated and heterogeneous mobility and the effect on
routing have recently been discussed in several
papers~\cite{cai08,bulut10,ciullo11,hossmann11}, but to our knowledge,
we are the first to provide a framework that accurately captures the
routing latency distribution for real traces with heterogeneous and
correlated movements.
\section{Conclusions}
\label{sec:conclusions}

Node mobility is often one of the most important factors to determine
the performance of wireless multi-hop communication. However, the
disparity of different application scenarios, and the difficulty of
obtaining representative large-scale mobility traces have often forced
the research community to use coarse and idealistic models for their
analyses. In this paper we proposed to model node connectivity as a
colouring process in the analysis of message dissemination
protocols. Such model of colouring times is strictly weaker in
assumptions on the contact patterns compared to the case where node
pairs are assumed to meet independently of other node pairs. Moreover,
this basic model of node connectivity can be used to derive the full
routing latency distribution of ideal epidemic routing as well as to
accurately approximate the routing latency distribution of multi-copy
routing. At a more general level, we believe that this approach could
also be valid for information dissemination in social computing
applications where individual actors interact in more complex
patterns.

Complex node mobility requires complex models of connectivity. When
modelling node connectivity as a set of inter-colouring distributions,
we need to collect more parameters from the mobility trace than just
the contact rate. Moreover, the calculations proposed in the paper
requires computer power to convolute the distributions. However, given
the complexity of the underlying systems, it is not surprising that
computer power is needed to provide good models. The key insight that
we try to convey is that node contact patterns should be modelled with
a sufficient level of fidelity to account for complex node
interactions. We believe that the colouring rate model for node
connectivity provides a good trade-off between abstraction and
fidelity.

\section{Acknowledgements}
This work was supported by the Swedish Research Council (VR) grant
2008-4667.

\appendix

\section{Proofs}

In this appendix we present a semi-formal\footnote{We try to avoid a
  too heavy use formal notation since it would require first
  presenting an extensive mathematical framework} proof of the
proposition that $\bot_C$ models a strictly wider set of contact point
processes than the $(\mathcal{S} \land \bot_{\le 1})$ model as
discussed in Section~\ref{sec:hierarchy}. We first show that if
$(\mathcal{S} \land \bot_{\le 1})$ is true for some point process, it
follows that $\bot_C$ is also true. We then proceed to show that the
opposite is not true.

\begin{theorem}
  $(\mathcal{S} \land \bot_{\le 1}) \Rightarrow \bot_C$
\end{theorem}
\begin{figure}[ht]
  \centering
  \includegraphics{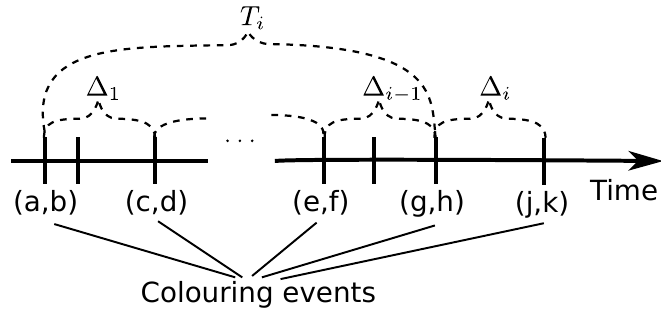}
  \caption{Inter-colouring times}
  \label{fig:inter-colouring}
\end{figure}
\begin{proof}
  Consider a set of nodes whose contact events constitute a
  multivariate point process for which $(\mathcal{S} \land \bot_{\le
    1})$ is true, meaning that the contact events for each pair of
  nodes is independent from every other pair of nodes, and the
  processes are stationary. Moreover, consider an arbitrary colouring
  process $C(t_0,s)$ and its associated sequence of inter-colouring
  times $\Delta_0,\Delta_1,\ldots$, we will now prove that for any $i$
  the sum $T_i = \sum_{j=1}^{i-1}\Delta_j$ is independent from the
  next inter-colouring time: $T_i \bot \Delta_i$.

  Figure~\ref{fig:inter-colouring} shows an arbitrary colouring
  process where $i$ nodes have been coloured. The time to colour one
  more node $\Delta_i$ will depend on all future meeting events
  $(x,y)$ such that either $x$ or $y$ has been previously coloured but
  not both. (In the particular case of the figure the first such
  contact event is between nodes $j$ and $k$.) Note that since at
  least one of the nodes in all relevant contact events $(x,y)$ one
  node is not coloured whereas all previous colour events contain only
  coloured nodes, we know from the assumption $\bot_{\le 1}$ stated in
  section \ref{sec:hierarchy} that all point processes $P^{(x,y)}$ are
  independent from all point processes whose events are part of the
  colouring process up to the $i$th colouring. Since the event $(j,k)$
  thus is independent from all previous events in the colouring
  process (including meetings $(a,b)$ and $(g,h)$), then the time
  $\Delta_i$ will also be independent from the time from event $(a,b)$
  to $(g,h)$, so $T_i\bot \Delta_i$.

\end{proof}

\begin{theorem}
  $\bot_{C} \nRightarrow (\mathcal{S} \land \bot_{\le 1})$
\end{theorem}
\begin{figure}[ht]
  \centering
  \includegraphics{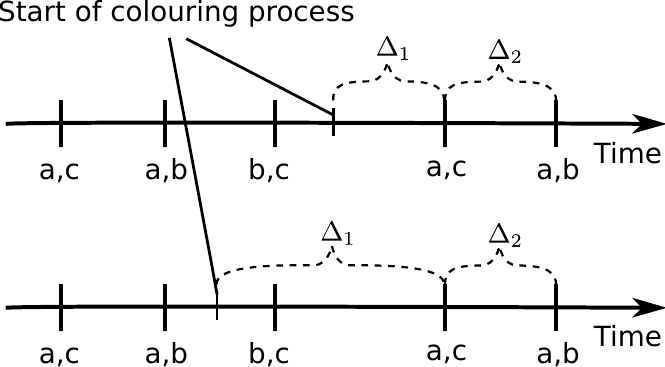}
  \caption{Counterexample}
  \label{fig:counterexample}
\end{figure}
\begin{proof}
  We will prove this with a counterexample of a contact process that
  fulfils $\bot_{C}$, but not $\bot_{\le 1}$. Consider a set of three
  nodes $a,b,c$. Assume that contacts always occur according to the
  following infinite sequence: \newline $\ldots, (a,b), (b,c), (c,a), (a,b),
  \ldots$. Moreover, assume that the time between two such contacts is
  independent and identically distributed with an exponential
  distribution. However, the \emph{pairwise} contacts are clearly not
  independent of each other since they are ordered, so $\bot_{\le 1}$
  is not true.

  What remains is to show that $\bot_{C}$ is true. Any colouring
  process in this system will have only two inter-colouring times:
  $\Delta_1$ and $\Delta_2$, so $T_1 = 0,\; T_2 = \Delta_1,\; T_3 =
  \Delta_1 + \Delta_2$. For $\bot_C$ to hold we need to show that $T_1
  \bot \Delta_1$ (follows immediately, since $T_1 = 0$) and $T_2 \bot
  \Delta_2$, which is the same as $\Delta_1 \bot \Delta_2$.

  The length of $\Delta_1$ depends on at what point in the contact
  sequence the colouring process starts and if the next meeting
  involves the only coloured node or not (see
  Figure~\ref{fig:counterexample}).  The next inter-colouring time
  $\Delta_2$ is always exactly equal to the time between the next two
  contacts in the system. By the assumptions of our counterexample
  this time is independent from all other inter-colouring times. Since
  $\Delta_1$ is only decided by the start time and the previous contact
  times it follows that $\Delta_1 \bot \Delta_2$, which means that
  $\bot_{C}$ holds. Thus, we have identified an example where $\bot_C$
  is true and $\bot_{\le 1}$ is false.

\end{proof}

\end{document}